\documentclass[twocolumn,showpacs,preprintnumbers,amsmath,amssymb]{revtex4}

\usepackage{subfigure}
\usepackage{graphicx}
\usepackage{dcolumn}
\usepackage{bm}
\usepackage{placeins}
\usepackage{verbatim}

\begin{document}

\preprint{APS/123-QED}

\title{A dynamical phase transition in ferromagnetic granular materials}

\author{Ran Itay}
 \author{Shlomo Havlin}%
 \author{Richard Berkovits}

\date{\today}

\begin{abstract}
	We study, using simulations the dynamical properties of complex ferromagnetic
	granular materials. The system of grains is modeled by a disordered
	two-dimensional lattice in which the grains are embedded, while 
	the magnitude and direction of the easy axis are random.
	Using the monte-carlo method we track the dynamics of the magnetic moments
	of the grains.
	We observe a transition of the system from a
	macroscopic blocked (ferromagnetic) phase at low 
	temperature in which the grain's magnetic 
	moment do not flip to the other direction to an unblocked (superparamagnetic) 
	phase at high temperature in which the magnetic moment is free to rotate.
	Our results suggest that this transition exhibits the characteristics
	of a second order phase transition such as the appearance of a giant cluster
	of unblocked grains which is fractal at the critical temperature,
	a peak in the size of the second largest cluster at the same temperature 
	and a power law distribution of cluster sizes near the criticality.
\end{abstract}

	 \pacs{64.60.aq, 64.60.ah, 45.70.vn, 75.30.kz}
\maketitle

\section{\label{sec:Introduction}Introduction:\protect\\}

\begin{figure}
	\includegraphics[width=0.25\textwidth]{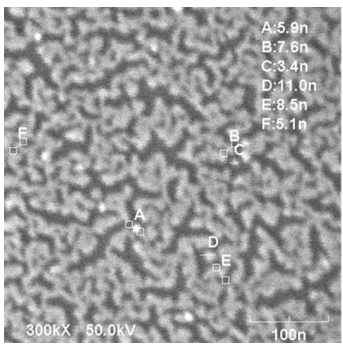}
	\caption{\label{fig:sem} SEM image of a granular material. A-F are single
	grains in the system, the average size is a few nanometers. The image was
	taken by CABL-9000C High Resolution Electron Beam Lithography System CRESTEC. }
	\end{figure}

	Granular systems composed of small metallic grains
	deposited on an insulating substrate (see Fig.~\ref{fig:sem}) 
	have generated much recent experimental and theoretical interest
	\cite{granular_alumin,tunel1,tunel2,granular1,thin_film,strelniker1,strelniker2,aviad1,aviad2}.
	These systems show a rich behavior stemming from the properties of the individual grains and
	from their mutual interaction leading to a complex collective behavior of the
	granular system. The size of the individual grain has an enormous influence on its 
	magnetic properties \cite{Cullity}. The grains may be classified into three categories
	according to their magnetic behavior: Multi-domain grains
	containing different magnetic domains (large
	grains), single-domain ferromagnetic grains, containing a single ferromagnetic domain (smaller
	grains) and superparamagnetic grains, in which the magnetic
	moment easily switches and therefore effectively have no average magnetic moment in the absence
	of an external field (ultra-small  grains). 
		
	Theoretical models describing single-domain
	grains were introduced in the 1930's
	\cite{frenkelDorfman,kittle,neel,stonerWohlfarth}. In 1950's 
	several experiments ~\cite{kittle} found that the typical size
	of these grains is a few hundred Angstroms. At low temperatures
	the direction of the magnetic moment is more or less fixed (up
	to small fluctuations around its preferred direction), but once
	temperature is high enough the magnitude of the magnetic moment 
	remains fixed but its direction becomes random. Thus for a 
	single domain uniaxial grain, as function
	of temperature, the grain changes from ferromagnetic, to a state
	in which it flips the magnetic moment up or down along its easy axis.
	The latter state was termed by Bean \cite{beanLivingston}
	as superparamagnetic, in which
	although at any given time the grain has a magnetic moment, on average
	it is zero ~\cite{magnetism,Cullity}.

	For a uniaxial grain
	the flipping rate is given by ~\cite{Vilchik}:
	\begin{equation}
	\label{eq:flip_rate}
	\frac{1}{\tau} = f_0e^{- \frac{KV}{k_BT}},
	\end{equation}
	where $f_0$ is the frequency factor, $k_BT$ is the thermal energy, V is the
	volume of the grain and K is an anisotropic constant (e.g. for shape anisotropy
	$K = \frac{\mu_0}{2}I^2_s$, where $\mu_0$  is the magnetic vacuum permeability and $I_S$ 
	is the saturation magnetization, i.e. the external field required
	to align the magnetic moment of the grain perpendicular to its easy axis).
		
\begin{figure}
		\includegraphics[width=0.25\textwidth]{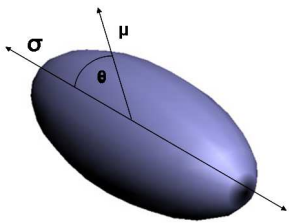}
		\caption{\label{fig:ellipsoid} An ellipsoid shaped grain, with $\mu$
		representing the direction of the dipole moment, $\sigma$ representing the
		direction of the easy axis along the semi-major axis of the ellipsoid and
		$\theta$ is the angle between them.}
	\end{figure}

	Considering the case where the grain has an ellipsoid like shape with semi-axis
	$a> b = c$, the semi major axis $a$ is called the easy axis (denoted
	$\hat{\sigma}$). Due to this geometric property there is a magnetic shape
	anisotropy ~\cite{Cullity}. 
	The Hamiltonian describing these grains is given by,~\cite{single_hamiltonain}
	\begin{equation}
	\label{eq:one_grain}
	H = \frac{\mu_0}{2}I^2_s\nu\sin^2 {\theta} - \mu_0 I_s \frac{\vec{H_0}\cdot
	\vec{\mu}}{|\vec{\mu}|},
	\end{equation}
	$\nu = N_a - N_b$ where $N_a$ and
	$N_b$ are the demagnetization coefficients of the ellipsoid along the a and b
	axes ($\nu =1$ for $N_a \gg N_b$ in MKS units), $\vec{\mu}$ is the magnetic
	moment of the grain, $\theta$ is the angle between $\vec{\mu}$ and
	$\hat{\sigma}$ (see Fig.~\ref{fig:ellipsoid}) and $H_0$ is an external
	magnetic field.
		
	For a system composed of many grains, magnetic interactions between the grains
	must be considered. Those take  the form of
	dipole-dipole interactions ~\cite{Vilchik}. The magnetic field of a magnetic
	moment $\mu$ belonging to grain $i$ at a distance $\vec{R}$ from the dipole is:
	\begin{equation}
	\label{eq:dipoleInteraction}
	\vec{B_i} =
	\frac{\mu_0}{4\pi}\frac{|\vec{\mu}|}{R^3}(2\cos{\phi}\hat{e}_R+\sin{\phi}\hat{e}_\phi),
	\end{equation}	
	where $\phi$ is the angle between $\vec{\mu}$, and $\vec{R}$, R is the
	magnitude of $\vec{R}$, $\hat{e}_R$ is a unit vector in the $\vec{R}$ direction and
	$\hat{e}_\phi$ is a unit vector in the $\phi$ direction (perpendicular to
	$\hat{e}_R$ on the $\vec{\mu} - \vec{R}$ plane).
	Summing up the contribution of all other grains to the magnetic field at
	the location of the $i$-th grain ($\vec{B}_{tot}$) results
	in the following Hamiltonian (see Eq.~\ref{eq:one_grain}) for the i-th
	grain:
	\begin{equation}
	\label{eq:final_Hamiltonian}
	H_i = \frac{\mu_0}{2}I^2_s\nu\sin^2 {\theta} - \mu_0 I_s
	\frac{(\vec{H}_0+\frac{\vec{B}_{tot}}{\mu_0})\cdot\vec{\mu}}{|\vec{\mu}|},
	\end{equation}
	where the system's Hamiltonian is the sum of the individual
	grain Hamiltonians: $H=\sum_i H_i$. Solving this Hamiltonian in temperature is
	a complex many-particle problem since the magnetic moment of each grain
	depends on the other grains moments.
		
	In this paper we study the transition
	from the blocked phase (where grains do not flip their magnetic moment) 
	at low temperatures to the unblocked phase (where the grains magnetic moment
	flip frequently) at higher temperatures. Our results suggest that 
	the crossover between those two phases follows the behavior expected from
	a macroscopic second order percolation type phase transition.
	This is shown by studying critical exponents and the fractal properties of the
	clusters of blocked/unblocked grains in the vicinity of the transition
	~\cite{intro2PercolationTheory,fractalsNdisorder}.

\section{\label{sec:Model}The Model:\protect\\}	
	
\begin{figure}
		\includegraphics[width=0.5\textwidth]{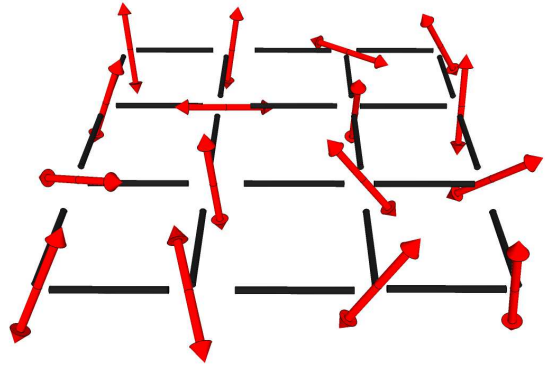}
		\caption{\label{fig:the_net} A 2D Square  lattice where the arrow location
		represents the location of a grain in the system (only on the x-y plane) and
		the direction of the arrow represents the direction of the easy axis
		of each grain (which may point at any random direction in the 3-dimensional
		space).}
	\end{figure}

	We consider ellipsoid grains which have an easy-axis along the semi-major 
	axis (see Fig.~\ref{fig:ellipsoid}). These grains are arranged in our 
	model on a two dimension square lattice, where
	each site represents a grain. Each grain has a dipole moment which may take 
	any direction in the three dimensional space and a randomly chosen easy axis 
	(which may take any direction, see Fig.~\ref{fig:the_net}). 
	Disorder is added to the system by defining a random distance between the
	sites, composed of the Euclidean distance between the sites added
	to a random component drawn from a Gaussian distribution of variance $0.3 a$ 
	(where $a$ is the lattice constant). The direction of the easy axis
	is chosen randomly with an equal probability at any direction while its
	magnitude is chosen from a Gaussian distribution with an average 
	$\langle I \rangle = 2$ and variance $0.3$.
	
	In order to simulate the behavior of the magnetic moments of the grains
	we use the Metropolis Monte-Carlo algorithm.  At each iteration all the grains are passed 
	sequentially. For each grain a new random 
	magnetic moment direction is chosen, then the total energy of the system is calculated 
	using the Hamiltonian (Eq.~\ref{eq:final_Hamiltonian}). In cases that the energy
	difference between the two configuration of the systems (i.e. before and after
	changing the direction of the magnetic moment of the grain) is negative we
	accept the change, however if the energy difference is positive we accept it
	with probability of $p= e^{-\beta\Delta E}$. When the magnetic moment direction
	of the grain crosses the plane perpendicular to its easy axis it is denoted as
	a flip. We track the number of flips each grain undergoes, then
	by dividing it by the total number of iterations (i.e., time steps) we obtain
	the probability of the magnetic moment to flip its direction for each time step. 
	A grain that has zero probability to flip is denoted as blocked while a grain
	that has a non-zero probability is regarded as unblocked. The system is then
	divided into clusters of blocked and unblocked grains, where
	a connected component that all of its sites are blocked is denoted as a blocked
	cluster. We preform this process for networks with the following sizes: 900,1600
	and 2500 grains arranged on a 30X30, 40X40, and 50X50 lattices. In
	order to get good statistics we run the simulations over 285,169 and
	100 realizations respectively. For each simulation we preform 120 iterations
	at every temperature. One can think that the monte-carlo algorithm is
	simple and fast, and can work on systems larger then a 50X50 lattice.
	However, due to these repetitions explained above this is not the case
	and on top of the standard monte-carlo algorithm complexity which is
	$n^4$ (n is the number of nodes on one side of the lattice(30,40,50)) one
	should take into account the number of realizations, the number of
	temperatures, and the number of iterations. This leads to a higher complexity
	which prevented us to run the simulations on a larger scale.
	  
	We explore the behavior of the system by using methods borrowed from
	complex network and percolation theory ~\cite{Barabási15101999,Newman2006,
	Dorogovtsev2003, pastor2004evolution,
	Barthelemy2011,cohen_1,
	song, neurochip, secondLargestCluster, intro2PercolationTheory}. The
	probability to perform a moment flip at a given time step is considered as the
	percolation probability. We then consider properties such as the size of the
	largest and second largest blocked cluster, the clusters size distribution
	and their fractal dimension in order to characterize the system and to suggest that
	there is a second order phase transition in the system with temperature as the
	system is crossing from a macroscopic unblocked phase at high temperature to a
	macroscopic blocked one at low temperature ~\cite{fractalsNdisorder,secondLargestCluster}.

\section{\label{sec:Results}Results:\protect\\}	
	
	\subsection{\label{subSec:prob2flip}Probability to Flip}
			
\begin{figure}
		\includegraphics[width=0.5\textwidth]{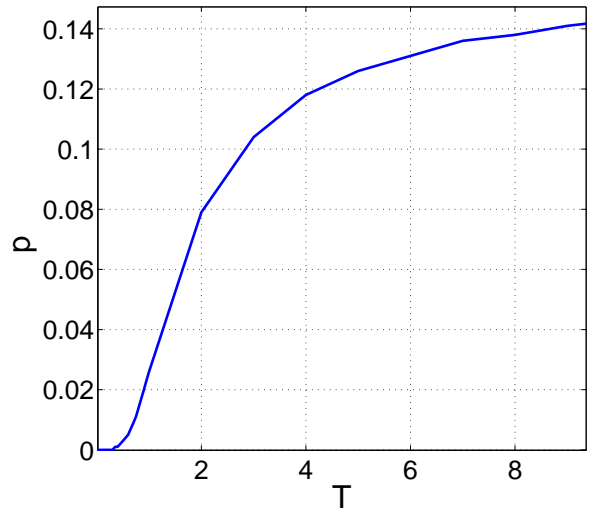}
		\caption{\label{fig:prob2flip}The probability to flip $p$ averaged over the
		entire system and on the ensemble of realizations as a function of temperature
		T. Our results suggest that the system undergoes a second order phase transition behavior.}
		\end{figure}

		The probability for a single grain to flip (Eq.~\ref{eq:flip_rate}) 
		decays exponentially with $\beta=\frac{1}{K_BT}$, however averaging the
		probability of a grain to flip over the entire system, when interactions are
		included, the temperature dependence is different. In
		Fig.~\ref{fig:prob2flip} we present the numerically averaged probability $p$
		for a single magnetic moment to flip, averaged over an ensemble of different
		realizations of grains as a function of temperature. By dividing the number
		of flips a grain performs by the number of iterations in each temperature
		(see Sec.~\ref{sec:Model}) we get the probability of a magnetic moment to flip
		at each time step. We verify this probability by changing the number of
		iterations to various numbers in the range 120-720, in all cases dividing
		the number of flips by the number of iterations gives the same probability.
		This behavior hints that the interactions between the magnetic moments of the
		grains qualitatively changes their collective behavior.

	\subsection{\label{subSec:Percolation}Percolation of Blocked Grains}
	
\begin{figure}[ht]
			\centering
			\subfigure[$T > T_c$]{
				\includegraphics[scale=0.37]{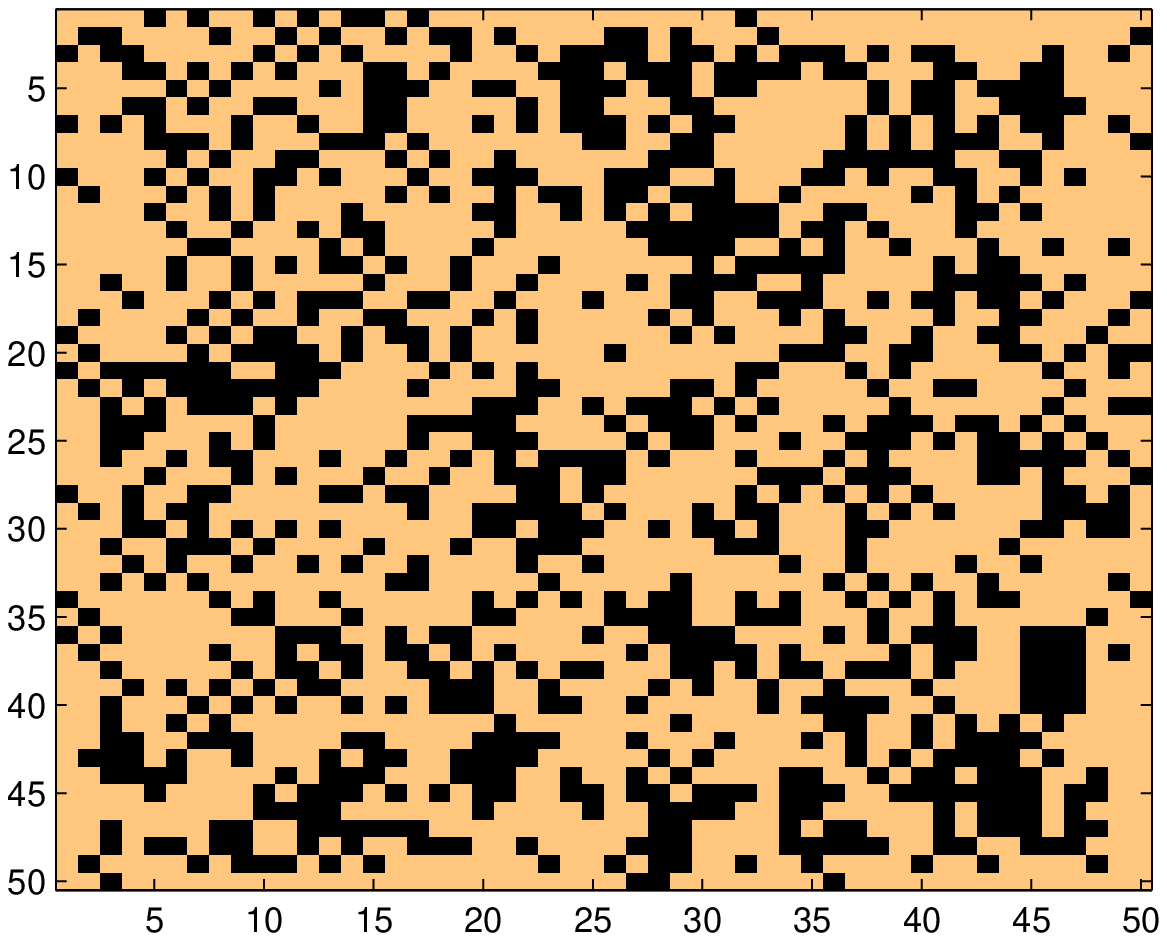}
				\label{fig:highT}
			}
			\subfigure[$T = T_c$]{
				\includegraphics[scale=0.37]{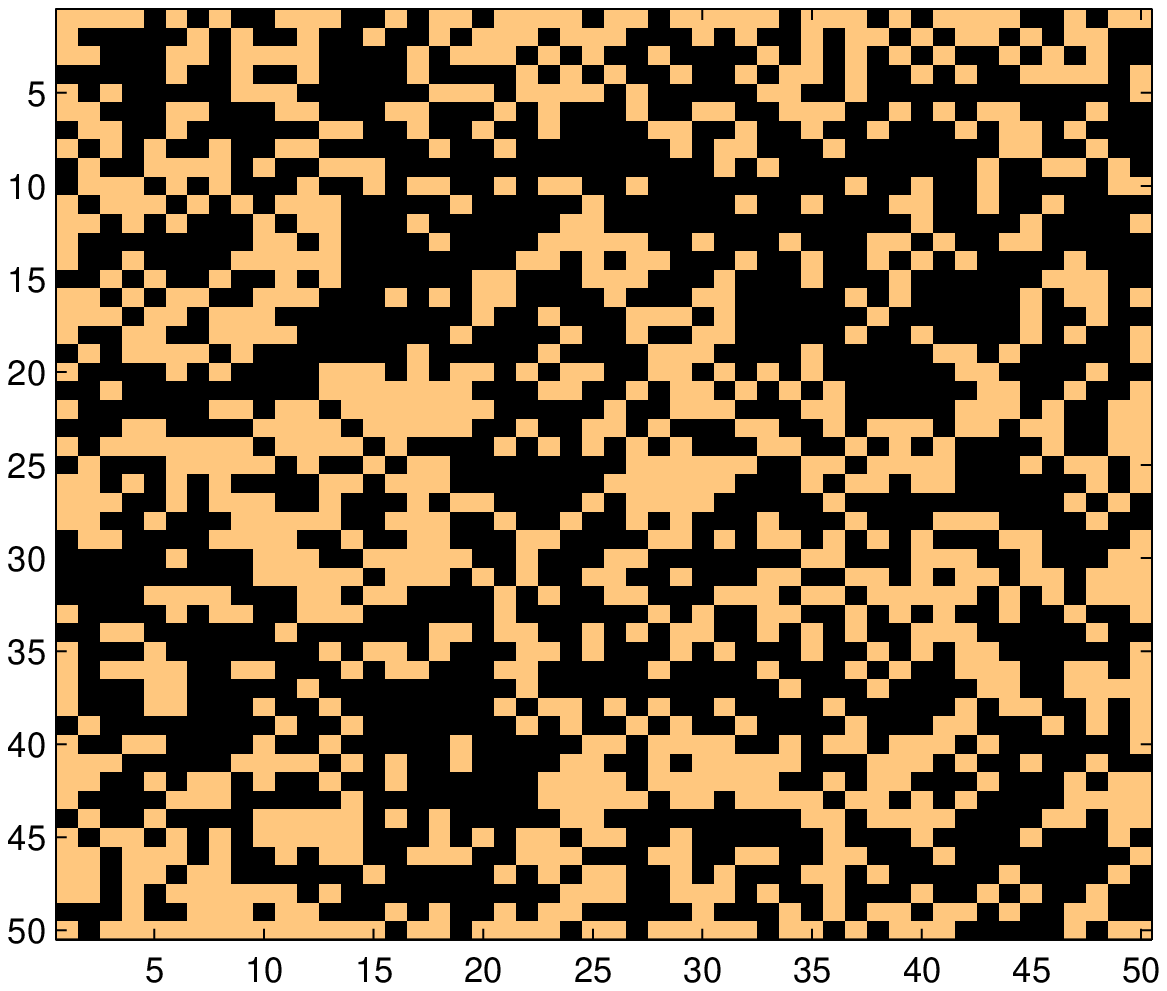}
				\label{fig:Tc}
			}
			
			\subfigure[$T < T_c$]{ 
				\includegraphics[scale=0.37]{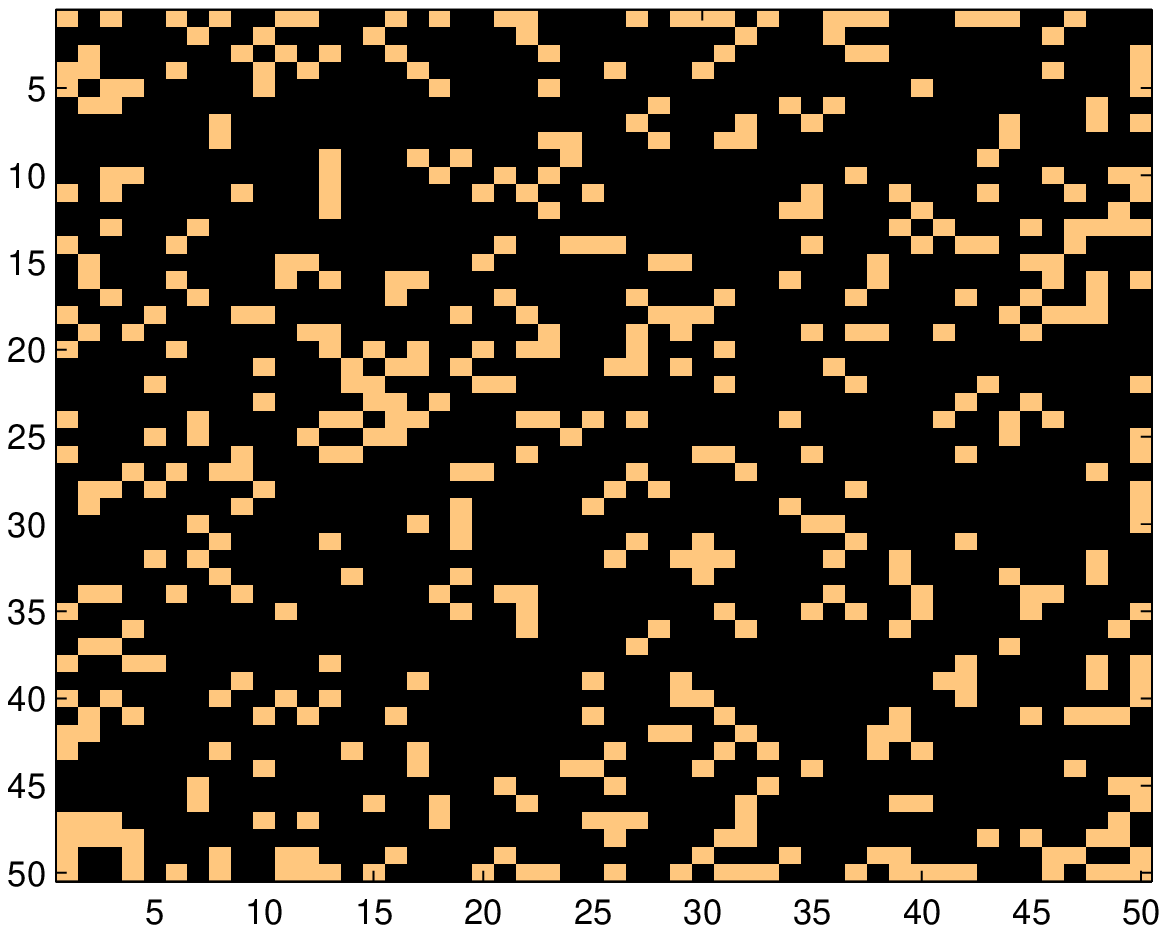}
				\label{fig:lowT}
				}
			\caption{\label{fig:multiColor}The grains location on the network at
			different temperatures. Black pixels represents blocked grains and light
			pixels represents unblocked ones. (a), $T>T_C$. (b) $T=T_c$.(c) $T<T_c$.}
		\end{figure}

		In order to deepen our understanding of the system's behavior 
		we plot in Fig.~\ref{fig:multiColor} the spatial distribution of blocked
		grains. The results are analogous to percolation theory where
		the temperature $T$ in our model plays the role of $q$, the chance of having
		an empty site, in the percolation model. We
		notice that above a given temperature , $T_c$, most of the grains are
		unblocked and there are only some small blocked clusters in the system.
		However below $T_c$ we notice that the system is dominated by blocked grains
		and only small unblocked clusters remain. Finally, at $T=T_c$ we notice that
		an almost equal number of blocked and unblocked grains, intermingled in very
		large clusters. This type of behavior is
		typical for systems close to a second order phase transition.
								
\begin{figure}
		\includegraphics[width=0.4\textwidth]{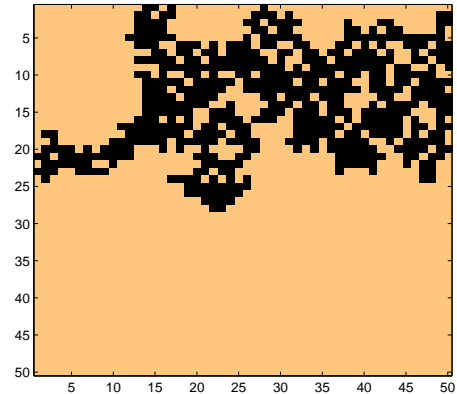}
		\caption{\label{fig:giantFractal}In black pixels, is shown the largest cluster
		in the system at $T = T_c$. Notice it's fractal shape.}
		\end{figure}

		Moreover observing the largest cluster (see Fig.~\ref{fig:giantFractal}) in
		the system at the critical point $T = T_c$ which is a reminiscent of a fractal
		nature, suggesting the divergence of the correlation length, a typical
		behavior of a second order phase transition.
		
\begin{figure}
		\includegraphics[width=0.5\textwidth]{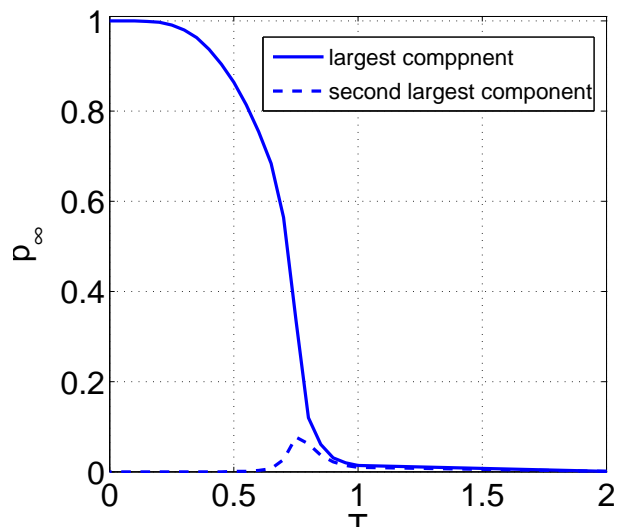}
		\caption{\label{fig:giantSecond}The size of the two largest components as a
		function of temperature. The full line represents the fraction of the
		network that is connected to the giant component , and the dashed line
		represents the fraction connected to the second largest cluster}
		\end{figure}

		Upon measuring the sizes of the two largest blocked clusters as a function of
		temperature (see Fig.~\ref{fig:giantSecond}) the analogy to percolation is
		reinforced. In percolation theory, the largest cluster (called the
		giant component) is zero for high values of $q$ and approaches the system's
		size when $q$ is zero \cite{fractalsNdisorder,intro2PercolationTheory}. The
		size of the second largest cluster approaches zero for high and  low $q$
		values, however when approaching the critical percolation threshold the size
		of the second largest cluster has a sharp maxima at $q_c$, the critical
		percolation value. By running a BFS (Breadth First Search) algorithm
		\cite{cormen} on the simulations results, the blocked clusters of the system
		are identified and their size is measured. The blocked clusters behave in a
		similar way to percolation clusters. The giant component's size grows to the
		system's size as $T$ approaches zero, and the size of the second largest
		cluster is zero for high and low temperatures and has a maximal value at
		$T_c$. In order to test the dependence on the system's size we measure these
		sizes on a scaled unit $p_{\infty}$ which is the number of connected nodes in
		the giant component divided by the number of nodes in the system. (see
		Fig.~\ref{fig:giantSecond}).

\begin{figure}
		\includegraphics[width=0.5\textwidth , height=0.3\textheight]{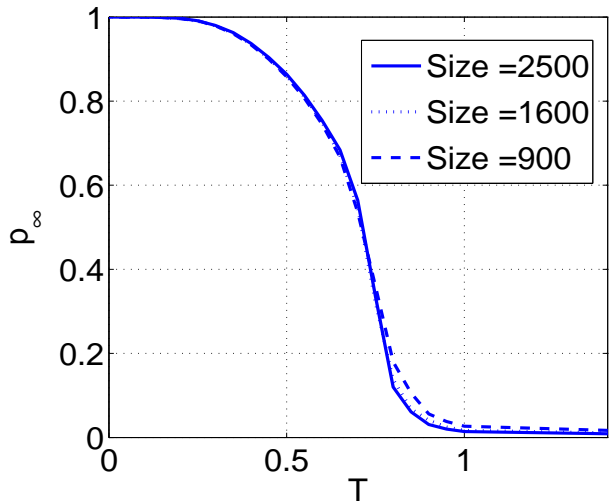}
		\caption{\label{fig:giantMulti}The fraction of the system covered by the giant
		component as a function of temperature for different size systems.}
		\end{figure}

		Thus, the properties of the giant and the second largest cluster seem to
		follow those expected from a second order phase transition.		
		Moreover, in order to verify that these properties are not size-dependent, we
		analyzed the giant component of systems with different sizes. We see that all
		system sizes show the same behavior and the giant component emerges at
		the same temperature. We also note that the transition becomes sharper for
		larger systems which insinuates that the width around $T_c$ is due to finite
		size effects and for infinite size system the order parameter $p_{\infty}$ is
		expected to have a discontinuity in the first derivative (see
		Fig.~\ref{fig:giantMulti}).

\begin{figure}
		\includegraphics[width=0.5\textwidth ,
												height=0.25\textheight]{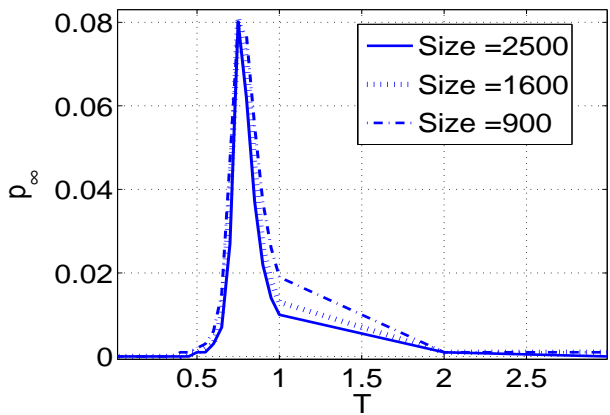}
		\caption{\label{fig:secondMulti} The fraction of the system covered the second
		largest cluster as a function of temperature for different size systems. Note
		that the peak for all systems is at the same temperature and at almost the
		same size.}
		\end{figure}

		The second largest cluster peak center also dose not change with size for
		our systems, and the fraction of the system covered by this cluster
		at $T_c$ is also similar for all system sizes but become sharper as the
		size of the system increases (see Fig.~\ref{fig:secondMulti}). Again this is
		in agreement with expectations a second order phase transition.
	
	\subsection{\label{subSec:critical_exponents}Critical Exponents}

\begin{figure}
		\includegraphics[width=0.5\textwidth,height=0.25\textheight]{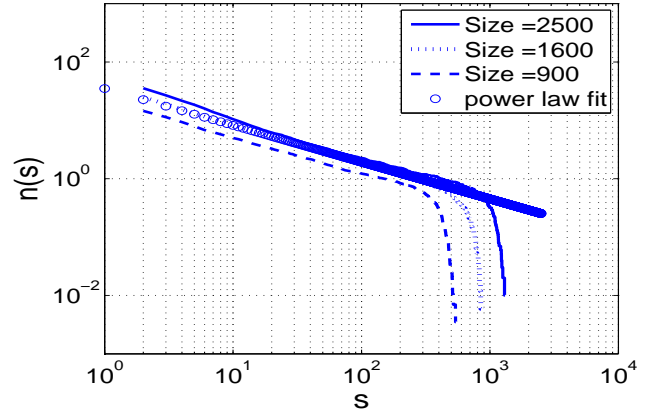}
		\caption{\label{fig:multiDist}The cumulative cluster size distribution at
		$T_c$. The cumulative distribution of the clusters sizes obeys a power law
		distribution.}
		\end{figure}

		Examining the results described above (see Sec.~\ref{subSec:Percolation})
		one can quantify the transition properties by analyzing
		the scaling properties of the blocked clusters. 
		The emergence of scaling laws is
		expected from systems undergoing a second order phase
		transition in which the correlation length diverges.
		The size distribution of blocked clusters can be used as indication for a 
		transition, since at transition, there should be no typical cluster
		size, i.e., the distribution should follow a power-law.
		Indeed, examining the blocked cluster size distribution in our system strongly
		suggests the existence of a scaling law. As shown in
		Fig.~\ref{fig:multiDist} the cluster size distribution $n(s)$, is in fact
		a power-law $n(s) \sim s^\tau$, typical for second order phase transitions.
		Moreover, we can see the same slope for systems with different sizes,
	 	indicating that they all converge to the same critical exponent ($\tau =
	 	1.63 $). In addition, as we increase the system size the cut-off value increases as
		well, thus one may expect that for an infinite size system there will be
		no cut-off (see Fig.~\ref{fig:multiDist}).

		In systems that undergo a second order phase transition it is known that the
		correlation length diverges at the critical point. The divergence of the
		cutoff in the cluster size distribution seen in Fig~\ref{fig:multiDist} is
		a signature of a diverging correlation length. The fractal nature of the
		clusters is another way of showing this divergence, as is well known both for
		percolation and Ising systems ~\cite{Ising}. At the critical point voids with
		no characteristic length scales appear in the giant component,  hence the
		correlation length which is the typical size of these voids diverges (see
		Fig~\ref{fig:giantFractal} ). This may be demonstrated by calculating the
		fractal dimension of the clusters.
				
\begin{figure}
		\includegraphics[width=0.5\textwidth]{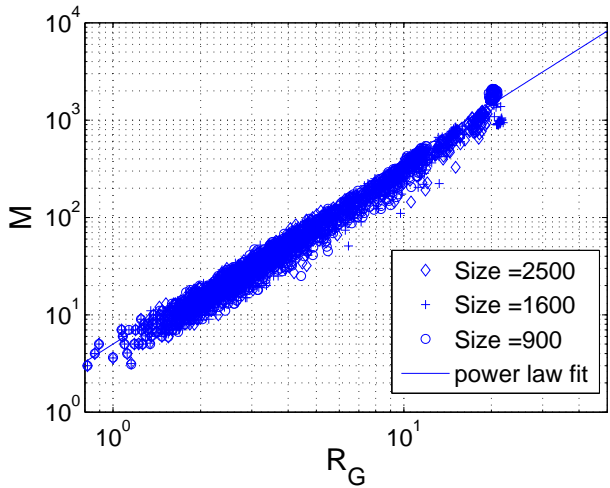}
		\caption{\label{fig:fracDim} The mass of the clusters as a function their
		radius of gyration at criticality. The
		slope represents the fractal dimension of the blocked clusters in the systems,
		$d_f = 1.89$, which is very close to that of percolation (1.896).}
		\end{figure}

		In order to calculate the fractal dimension (noted $d_f$) of the
		blocked clusters in the system, we use the following relation
		\cite{Percolation,secondLargestCluster,song,fractalsNdisorder}:
		\begin{equation}
		\label{eq:fractal}
		M \sim R_G^{d_f},
		\end{equation}
		where $M$ is the mass of the cluster and 
		$R_G=\sqrt{\langle R-R_{cm} \rangle}$
		is the radius of gyration ($R_{cm}$ is the radius of the center of mass). As
		we have shown above (see Fig.~\ref{fig:giantFractal}) the cluster in our
		system has a fractal nature. By plotting the mass of the clusters as a
		function of their $R_G$ on a log-log scale, we can see they are obeying  a
		power low (see Fig.~\ref{fig:fracDim}). The slope of the power low fit is the
		dimension of the clusters which are fractal and their fractal dimension is 
		$d_f \approx 1.89$. Systems with different sizes have the same fractal
	    dimension.
			
\section{\label{sec:Summary}Summary:\protect\\}	

	In summary, by using the analogy between the behavior of
	blocked/unblocked grains in magnetic granular systems composed of nanometric 
	ferromagnetic grains and percolation,
	we suggest that the granular system exhibits a new kind of
	second order phase transition. In this transition the role of the 
	order parameter is played by the dynamics of the magnetic moment of
	the grain, i.e. ,whether the magnetic moment of the grain can flip or it is
	blocked. The second giant component has a peak at the critical temperature
	$T_c$, the cluster size distribution obeys a power law, and the clusters have
	a fractal dimension. All these properties can be seen for systems of
	different sizes which support the assumption that the transition we observe is
	a real thermodynamic second order phase transition.
		
	From the experimental point of view, we believe that this transition,
	in principal, could be observed by direct measurements of the local magnetic
	field using a SQUID (superconducting quantum interference device), 
	or an MFM (Magnetic force microscope) tip. Another method, which is probably
	easier from the experimental point of view, is to identify the transition
	by its effects on the electrical transport through these
	systems ~\cite{strelniker1,strelniker2}, especially the resistance noise. We
	expect that the transport noise will be strongly enhanced in the vicinity of the transition due to 
	the divergence of the correlation length. 
	
\begin{acknowledgements}
		We would like to thank Avi Gozolchiani, Aviad Frydman and Yakov Strelniker for
		helpful and fruitful discussions. We also wish to thank Tal Havdala for
		providing the CRESTEC image(Fig.~\ref{fig:sem}).We
		thank the European EPIWORK project, the Israel Science Foundation, ONR, DFG,
		and DTRA for financial support.
\end{acknowledgements}

\end{document}